\documentclass[14pt]{article}
\usepackage{amsmath}
\usepackage{amssymb}
\usepackage{amsfonts}
\usepackage{amscd}
\usepackage{delarray}
\usepackage{graphicx}

\newcommand{\R}{\mathbb{R}}

\begin{document}

\title{A formula for the spectral projection of the time operator}
\author{M. Courbage and  S. M. Saberi Fathi}
\date{ }

\maketitle \centerline{\it  Laboratoire Mati\`ere et Syst\`emes
Complexes (MSC) } \centerline{\it UMR 7057 CNRS et Universit\'e
Paris 7- Denis Diderot } \centerline{\it Case 7056, B\^{a}timent
Condorcet, 10, rue Alice Domon et Léonie Duquet} \centerline{\it
75205 Paris Cedex 13 / FRANCE} \centerline{\it emails :
courbage@ccr.jussieu.fr, saberi@ccr.jussieu.fr} \vspace{1cm}

{\footnotesize\textbf{Abstract}: In this paper, we study the
one-level Friedrichs model with using the quantum time
super-operator that predicts the excited state decay inside the
continuum. Its survival probability in long time limit is an
algebraically decreasing function and an exponentially decreasing
multiplied by the oscillating functions.}

\bigskip

\section{Introduction}
In this paper we shall study the concept of survival probability of
an unstable quantum system introduced in \cite{courbage} and we
shall test it in the Friedrichs model \cite{fried}. The survival
probability should be a monotonically decreasing time function  and
this property could not exist in the framework of the usual
Weisskopf-Wigner approach \cite{misrasud,horwitz,misrasudchiu,bohm}.
It could only be properly treated if it is defined through an
observable time operator $T$ whose eigenprojections provide the
probability distribution of  the time of decay. The equation
defining time operator is the following:
\begin{equation}
U_{-t} T U_{t} = T + tI
\end{equation} where $ U_{t}$ is the unitary group of states
evolution. It is known that such an operator cannot exist when the
evolution is governed by the Schr\"{o}dinger equation, since the
Hamiltonian  has a bounded spectrum from below, and this contradicts
the equation:
\begin{equation}
[H , T] = iI
\end{equation}
in the Hilbert space of pure states $\mathcal{H}$. However, a time
operator can exist under some conditions, for mixed states.  They
can be embedded \cite{courbage,cour1,mpc} in The ``Liouville space",
denoted $ \mathcal{L}$, that is the space of Hilbert-Schmidt
operators $\rho $ on $ \mathcal{H}$ such that $Tr(\rho^* \rho) <
\infty$,   equipped with the scalar product: $<\rho, \rho'> =
Tr(\rho^* \rho ') $. The time evolution of these operators is given
by the  Liouville von-Neumann group of operators:
\begin{equation}
U_t\rho = e^{-itH}\rho e^{itH}
\end{equation}
The infinitesimal self-adjoint generator of this group is the
Liouville von-Neumann operator $L$ given by:
\begin{equation}
L\rho = H\rho - \rho H
\end{equation}
That is, $U_t = e^{-itL}$.  The states of a quantum system are
defined by  normalized elements $\rho \in \mathcal{L}$ with respect
to the scalar product, the expectation of $T$ in the state $\rho$ is
given by:
\begin{equation}
<  T > _\rho = < \rho, T\rho>
\end{equation}
and the ``uncertainty" of the observable $T$ as its fluctuation in
the state $\rho$:
\begin{equation}\label{tio13}
(\Delta T)_\rho = \sqrt{<  T^2 > _\rho - (<  T > _\rho)^2}
\end{equation}
Let $ \mathcal{P}_\tau$ denote the family  of spectral projection
operators of $T$ defined by:
\begin{equation}
T = \int_{\R} \tau d\mathcal{P}_{\tau}
\end{equation}
It is shown that \cite{courbage} the unstable states are those
states verifying $\rho=\mathcal{P}_0\rho$. Let $\bigtriangleup E$ be
the usual energy uncertainty in the state $M$ given by:
\begin{equation}\label{tio16}
 \bigtriangleup E=\sqrt{\mathrm{Tr}(M.H^2)-(\mathrm{Tr}(M.H))^2}
\end{equation}
and $\bigtriangleup T = (\bigtriangleup T)_{M^{1/2}}$ be the
uncertainty of $T$ in the state $M$ defined as in (\ref{tio13}). It
has been shown that:
\begin{equation}\label{tio17}
\bigtriangleup E\bigtriangleup T\geq\frac{1}{2\sqrt{2}}
\end{equation}
This uncertainty relation leads to the interpretation of $T$ as the
time occurrence of specified random events. The time of occurrence
of such events fluctuates and we speak of the probability of its
occurrence in a time interval $I =]t_1, t_2]$. The observable $T'$
associated to such event in the initial state $\rho_0$ has to be
related to the time parameter t by:
\begin{equation}\label{tio18}
\langle T'\rangle_{\rho_t}=\langle T'\rangle_{\rho_0}-t
\end{equation}
where $\rho_t=e^{-itL}\rho_0$. Comparing this condition with the
above Weyl relation we see that we have to define $T'$ as: $T'=-T$.
Let $\mathcal{Q}_\tau$ be the family of spectral projections of
$T'$, then, in the state $\rho$, the probability of occurrence of
the event in a time interval $I$ is given, as in the usual von
Neumann formulation, by:
\begin{equation}\label{tio18-1}
\mathcal{P}(I,\rho)=\|\mathcal{Q}_{t_2}\rho\|^2-\|\mathcal{Q}_{t_1}\rho\|^2=
\|(\mathcal{Q}_{t_2}-\mathcal{Q}_{t_1})\rho\|^2:=\|\mathcal{Q}(I)\rho\|^2
\end{equation}
The unstable ``undecayed" states prepared at $t_0=0$ are the states
$\rho$ such that $\mathcal{P}(I, \rho) = 0$ for any negative time
interval $I$, that is:
\begin{equation}\label{tio19}
\|\mathcal{Q}_{\tau}\rho\|^2=0, ~~\forall\tau\leq0
\end{equation}
In other words, these are the states verifying $\mathcal{Q}_0\rho =
0$. It is straightforwardly checked that the spectral projections
$\mathcal{Q}_{\tau}$ are related to the spectral projections
$\mathcal{P}_{\tau}$ by the following relation:
\begin{equation}\label{tio20}
\mathcal{Q}_{\tau} =1-\mathcal{P}_{-\tau}
\end{equation}
Thus, the unstable states are those states verifying: $\rho =
\mathcal{P}o_0\rho$ and they coincide with our subspace
$\mathcal{F}_0$\footnote{We define the subspace $\mathcal{F}_{t_0}$
to the set of decaying states prepared at time $t_0$}. For these
states, the probability that a system prepared in the undecayed
state $\rho$ is found to decay sometime during the interval $I =]0,
t]$ is $\|\mathcal{Q}_{t}\rho\|^2=1-\|\mathcal{P}_{-t}\rho\|^2$ a
monotonically nondecreasing quantity which converges to $1$ as
$t\rightarrow\infty$ for $\|\mathcal{P}_{-t}\rho\|^2$ tends
monotonically to zero. As noticed by Misra and Sudarshan
\cite{misrasud}, such quantity could not exist in the usual quantum
mechanical treatment of the decay processes and could not be related
to the ``survival probability" for it is not a monotonically
decreasing quantity in the Hilbert space formulation. In the
Liouville space, given any initial state $\rho$, its survival
probability in the unstable space is given by:
\begin{equation}\label{tio21}
p_\rho(t)=\|\mathcal{P}_{0}e^{-\mathrm{i}Lt}\rho\|^2
\end{equation}
Hence, in the Liouville space, given any initial state $\rho$, its
survival probability in the unstable space is given by:
\begin{eqnarray}\label{115}
\nonumber p_{\rho}(t) &=& \| \mathcal{P}_0e^{-itL}\rho \|^2\\
\nonumber &=&\| U_{-t}\mathcal{P}_0U_t\rho \|^2\\
&=& \| \mathcal{P}_{-t}\rho \|^2
\end{eqnarray}
Then, the survival probability  is monotonically decreasing to $0$
as $t \rightarrow \infty$. As $\mathcal{P}_t$ is a spectral family
of projections $p_\rho(t)\rightarrow 1$ when $t\rightarrow-\infty$.
This survival probability and the probability of finding the system
to decay sometime during the interval $I = ]0, t],~
q_\rho(t)=\|\mathcal{Q}_\rho(t)\|^2$ are related by:
\begin{equation}
q_{\rho}(t)= 1 - p_{\rho}(t)\label{sp}
\end{equation}


\section{Spectral projections of time operator}
The expression of time operator is given in a spectral
representation of $H$. As shown in \cite{courbage}, $H$ should have
an unbounded absolutely continuous spectrum. In the simplest case,
we shall suppose that $H$ is represented as the multiplication
operator on ${\cal H}=L^{2}(\mathbb{R}^+)$ :
\begin{equation}
H\psi(\lambda)=\lambda\psi(\lambda)
\end{equation}
the Hilbert-Schmidt operators on $L^{2}(\mathbb{R}^+)$ correspond
to the square-integrable functions $ \rho(\lambda,\lambda^{'})\in
L^{2}(\mathbb{R}^+ \times \mathbb{R}^+ )$ and the Liouville-Von
Neumann operator $L$ is given by :
\begin{equation}
L \rho(\lambda,\lambda^{'})=
(\lambda-\lambda^{'})\rho(\lambda,\lambda^{'})
\end{equation}
Then we obtain a spectral representation of $L$ via the change of
variables:
\begin{equation}
\nu = \lambda-\lambda^{'}
\end{equation}
and
\begin{equation}
E= \min(\lambda,\lambda^{'})
\end{equation}
This gives a spectral representation of
 $L$:
\begin{equation}
L\rho(\nu, E)= \nu \rho(\nu, E),
\end{equation}
where $ L\rho(\nu,E)\in L^{2}(\mathbb{R} \times \mathbb{R}^+ )$. In
this representation $ T \rho(\nu, E)= \mathrm{i}
\frac{d}{d\nu}\rho(\nu, E)$ so that the spectral representation of
$T$ is obtained by the inverse Fourier transform:
\begin{equation}
\hat{\rho}(\tau, E) = \frac{1}{\sqrt{2 \pi}}
\int_{-\infty}^{+\infty} e^{\mathrm{i}\tau \nu}\rho(\nu, E)d\nu =(
{\cal F}^* \rho)(\tau, E)
\end{equation}
and
\begin{equation}
T\hat{\rho}(\tau, E) = \tau \hat{\rho}(\tau, E).
\end{equation}
The spectral projection operators ${\cal P}_s$ of T are given in
the $(\tau, E)$-representation by:
\begin{equation}
{\cal P}_s\hat{\rho}(\tau, E)=
\chi_{]-\infty,s]}(\tau)\hat{\rho}(\tau, E)
\end{equation}
where $\chi_{]-\infty,s]}$ is the characteristic function of
$]-\infty,s]$. So that we obtain in the $(\nu,E)$-representation
the following expression of these spectral projection operators:
\begin{eqnarray}
\nonumber{\cal P}_s\hat{\rho}(\nu, E)&=&
\frac{1}{\sqrt{2\pi}}\int_{-\infty}^{s}  e^{-\mathrm{i} \nu
\tau}\hat\rho(\tau, E)\,d\tau\\&=&e^{-\mathrm{i}\nu
s}\int_{-\infty}^{0} e^{-i\nu\tau}\hat\rho(\tau+s, E)\,d\tau.
\end{eqnarray}
Let us denote the Fourier transform ${\cal
F}f(\nu)=\frac{1}{\sqrt{2\pi}}\int_{-\infty}^{\infty} e^{-\mathrm{i}
\nu \tau}f(\tau)\,d\tau $ and remind the Paley-Wiener theorem which
says that a function $f(\nu)$ belongs to the Hardy class $H^+ $(i.e.
the limit as $y \rightarrow 0^+ $ of an analytic function
$\Phi(\nu+\mathrm{i} y)$ such that
$\int_{-\infty}^{\infty}\mid\Phi(\nu+\mathrm{i}\emph{y})\mid^2\,dy<\infty)$
if and only if it is of the form
$f(\nu)=\frac{1}{\sqrt{2\pi}}\int_{-\infty}^{0} e^{-\mathrm{i} \nu
\tau}\hat{f}(\tau)\,d\tau $ where $ \hat{f}\in L^2(\mathbb{R}^+) $
\cite{titch}. Using the Hilbert transformation:
\begin{equation}
Hf(x)=\frac{1}{\pi}\textsf{P}\int_{-\infty}^{\infty}\frac{f(t)}{t-x}\,dt\label{hilbert}
\end{equation}
 for $ f\in L^2(\mathbb{R}) $ we can write the decomposition:
\begin{eqnarray}
\nonumber f(x)&=&\frac{1}{2}[f(x)- \mathrm{i} H
f(x)]+\frac{1}{2}[f(x)+\mathrm{i} H f(x)]\\&=&f_+(x)+f_-(x)
\end{eqnarray}
According to the theorem, $f_+(x)$ (resp.$ f_-(x) $) belongs to
the Hardy class $ \emph{H}^+$( resp.$H^-$). This decomposition is
unique as a result of Paley-Wiener theorem. Thus taking the
Fourier transformation of $f$ we obtain :
\begin{eqnarray}
\nonumber{\cal F}(f)(\nu)=\frac{1}{\sqrt{2\pi}}\int_{-\infty}^{0}
e^{-\mathrm{i} \nu
\tau}\hat{f}(\tau)\,d\tau+\frac{1}{\sqrt{2\pi}}\int_0^{\infty}
e^{-\mathrm{i} \nu \tau}\hat{f}(\tau)\,d\tau.
\end{eqnarray}
It follows that:
\begin{equation}
\frac{1}{\sqrt{2\pi}}\int_{-\infty}^{0}  e^{-\mathrm{i} \nu
\tau}\hat{f}(\tau)\,d\tau=\frac{1}{2}({\cal F}(f)-\mathrm{i}
H{\cal F}(f)).
\end{equation}
Now, using the well known property of the translated Fourier
transformation $\sigma_s \hat{f}(\tau)=\hat{f}(\tau+s)$ we have :
\begin{equation}
{\cal F}(\sigma_s \hat{f})(\nu)=e^{\mathrm{i} \nu s}{\cal
F}.\hat{f}(\nu)= e^{\mathrm{i} \nu s}f(\nu),
\end{equation}
this and (\ref{hilbert}) yields:
\begin{equation}
{\cal P}_s\rho(\nu,E)=\frac{1}{2}e^{-\mathrm{i}\nu s}[e^{i\nu
s}\rho(\nu,E)-\mathrm{i} H(e^{\mathrm{i}\nu s}\rho(\nu,E))].
\end{equation}
Thus:
\begin{equation}
{\cal P}_s\rho(\nu,E)=\frac{1}{2}[\rho(\nu,E)-\mathrm{i}
e^{-\mathrm{i}\nu s}H(e^{\mathrm{i}\nu s}\rho(\nu,E))].\label{t1}
\end{equation}
It is clear from (\ref{115}) that ${\cal P}_s\rho(\nu,E)$ is in the
Hardy class $ H^+$ .


\section{Computation of spectral projections of \emph{T}
in a Friedrichs model}

The one-level Friedrichs model is a simple model Hamiltonian in
which a discrete eigenvalue the free Hamiltonian $H_0$. It has been
often used as a simple model of decay of unstable states
illustrating the Weisskopf-Wigner theory of decaying quantum
systems. The Hamilton operator $H$ is an operator on the Hilbert
space of the wave functions of the form $
\mid\psi>=\{f_0,g(\omega)\}, f_0\in\mathbb{C}, g\in
L^2(\mathbb{R}^+)$,
\begin{equation}
 H=H_0 + \lambda V,
\end{equation}
where $\lambda $ is a positive coupling constant, and
\begin{equation}
H_0 \mid \psi>=\{\omega_1 f_0,\omega g(\omega)\},   (\omega_1 >
0).
\end{equation}
We shall denote the eigenfunction of $H_0$ by $\chi = \{1,0\}$.
The operator \emph{V} is given by:
\begin{equation}
V\{f,g(\omega)\}=\{<v(\omega),g(\omega)>,f_0.v(\omega)\}.
\end{equation}
Thus $H$ can be represented as a matrix :
\begin{equation}
H=\begin{array}({cc}) \omega_1 & \lambda v^*(\omega)  \\ \lambda
v(\omega)
 & \omega
\end{array},
\end{equation}
where $v(\omega)\in L^2(\mathbb{R^+})$ and it is called a factor
form. It has been shown than for $\lambda$ small enough, $H$ has
no eigenvalues and that the spectrum of $H$ is continuous
extending over $\mathbb{R}^+$. It is also shown that in the
outgoing spectral representation of $H$, the vector $\chi$ is
represented by:
\begin{equation}
f_1(\omega)=\frac{\lambda v(\omega)}{\eta^+(\omega + \mathrm{i}
\epsilon)},
\end{equation}
where
\begin{equation}
\eta^+(\omega + \mathrm{i} \epsilon)=\omega - \omega_1 + \lambda^2
\lim_{\epsilon \to 0} \int_{0}^{\infty}
\frac{|v(\omega)|^2}{\omega^{'} - \omega - \mathrm{i }\epsilon} d
\omega^{'}
\end{equation}
and $H\chi$ is represented $\omega f_1(\omega)$. The quantity
$<\chi, e^{-\mathrm{i}Ht}\chi>$ is usually called the decay law
and $|<\chi, e^{-\mathrm{i}Ht}\chi>|^2= \int_{0}^{\infty}
|f_1(\omega)|^2 e^{-\mathrm{i} \omega t} d \omega $ is called the
survival probability at time $t$. It is however clear that this is
not a true probability, since it is not a mononically decreasing
quantity,  although it tends to zero as a result of the
Riemann-Lebesgue lemma. Let us now identify the state $\chi$ with
element $\rho= |\chi><\chi|$ of the Liouville space, that is, to
the kernel operator:
\begin{equation}
\rho_{11}(\omega, \omega^{'})=f_1(\omega)\overline{
f_1(\omega^{'}}).
\end{equation}
We shall compute first the unstable component $ {\cal P}_0
  \rho_{11}$ and show that $ {\cal P}_0
  \rho_{11}\neq \rho_{11}$. Then we shall compute the survival
probability in the state $\rho$.
\begin{equation}
\lim_{s\rightarrow \infty}\|{\cal P}_{-s} \rho \|^2  \rightarrow
0.
\end{equation}


\section{Computation of ${\cal P}_s \rho_{11}$}

As explained above the Liouville operator is given by:
\begin{equation}
L\rho(\omega, \omega^{'})=(\omega - \omega^{'})\rho(\omega,
\omega^{'})
\end{equation}
and that the spectral representation of $L$ is given by the change
of variables:
\begin{equation}
\nu=\omega - \omega^{'}
\end{equation}
and
\begin{equation}
E= \mathrm{min}(\omega, \omega^{'}).
\end{equation}
Thus we obtain for $\rho_{11}(\nu,E)$ :
\begin{equation}
\rho_{11}(\nu,E)=\left\{ \begin{array}{ll}
\lambda^{2}\frac{v(E)}{\eta^{-}(E)}\frac{v^*(E+\nu)}{\eta^{+}(E+\nu)}&
\mbox {$\nu >0$}\\\\
\lambda^{2}\frac{v^*(E)}{\eta^{+}(E)}\frac{v(E-\nu)}{\eta^{-}(E-\nu)}&
\mbox{$\nu < 0$}.\end{array} \right.
\end{equation}
where $\eta^-$ is the complex conjugate of $\eta^+$.
\begin{equation}
\eta^+(\omega) \simeq \omega - z_1,~~~~  z_1 =\widetilde{\omega}_1 -
\mathrm{i} \frac{\gamma}{2}
\end{equation}
where  $z_1$ is called the resonance with energy
$\widetilde{\omega}_1$ and a lifetime $\gamma$ \cite{marchand}. It
is believed that this form results from weak coupling
approximations. It can be shown $\rho_{11}(\nu,E)$ in the following
form:
\begin{equation}
\rho_{11}(\nu,E)=\frac{\gamma}{2}f(\nu),\label{eq}
\end{equation}
where
\begin{equation}
f(\nu)=\left\{ \begin{array}{ll}
 \frac{1}{\nu^*_0(\nu+\nu_0)}& \mbox{ $\nu > 0$}\\\\
\frac{1}{\nu_0(\nu^*_0-\nu)}& \mbox{ $\nu <0$}.
\end{array} \right.\label{c1}
 \end{equation}
where $\nu_0=a+\mathrm{i} b = (E-\widetilde{\omega}_1)+ \mathrm{i}
\frac{\gamma}{2}$. For obtaining ${\cal P}_s(f)(\nu)$, we shall
use the formula (\ref{t1}) and  we obtain
\begin{eqnarray} \nonumber\mathcal{P}_sf(\nu) =
\mathrm{i}e^{-\mathrm{i}s\nu}[\frac{-1}
{2\pi\nu_0(\nu_0^*-\nu)}(\int_{-\infty}^0
\frac{e^{-sy}}{y+\mathrm{i}\nu_0^*}dy- \int_{-\infty}^0
\frac{e^{-sy}}{y+\mathrm{i}\nu}dy)& &  \\
\nonumber +\frac{1} {2\pi\nu_0^*(\nu+\nu_0)}(\int_{-\infty}^0
\frac{e^{-sy}}{y-\mathrm{i}\nu_0}dy- \int_{-\infty}^0
\frac{e^{-sy}}{y+\mathrm{i}\nu}dy)]& & \\
+\left\{
\begin{array}{ll}
e^{-\mathrm{i}s\nu }[\frac{e^{\mathrm{i}s\nu_0^* }
}{\nu_0(\nu_0^*-\nu)}-\frac{e^{-\mathrm{i}s\nu_0 }
}{\nu_0^*(\nu_0+\nu)}],& E< \widetilde{\omega}_1 \\ \\
0, & E>\widetilde{\omega}_1.
\end{array} \right.\label{t5}
\end{eqnarray}
In this equation the non integrals terms yield a poles and lead to
the resonance shown in equation (\ref{res}), and the integral terms
yield an algebraical term analog to the background in the
Hamiltonian theories \cite{bohm,bohm1,bohmhar}. We can also compute
the same result for the case $\nu<0$.
\subsection{Case  $s=0$}
In this case (\ref{t5}) can be obtained as:
\begin{eqnarray}
\nonumber\mathcal{P}_0f(\nu) =\frac{\mathrm{i}}
{\nu_0(\nu_0^*-\nu)} \log^+(\frac{\nu}{\nu_0^*})
-\frac{\mathrm{i}}
{\nu_0^*(\nu+\nu_0)}\log^+(-\frac{\nu}{\nu_0}) \\
+\left\{
\begin{array}{ll}
[\frac{1}{\nu_0(\nu_0^*-\nu)}-\frac{1
}{\nu_0^*(\nu_0+\nu)}],& E< \widetilde{\omega}_1 \\ \\
0, & E>\widetilde{\omega}_1.
\end{array} \right.
\end{eqnarray}
where $\log^+z$ is the complex analytic function with  cut-line
along the negative axis:
\begin{equation}
\log^+z= \begin{array}{ll} \log|z| + \mathrm{i} \arg(z), &
\arg(z)\in ]-\frac{\pi}{2},\frac{3\pi}{2}[
\end{array}.
\end{equation}
Also, we used $ \lim_{R \to \infty}
{\log^+(\frac{\mathrm{i}\nu-R}{\mathrm{i}\nu_0^* - R}) \to 0}$~ and~
 $ \lim_{R \to \infty}
{\log^+(\frac{\mathrm{i}\nu-R}{-\mathrm{i}\nu_0 - R}) \to 0}$.

We see that $\mathcal{P}_0f(\nu)$ is an upper Hardy class function.
This verified the general theorem about the properties of the
unstable states associated to time operator, as being in the upper
Hardy class.


\subsection{Asymptotical behavior of the survival probability}
First, using the following approximation, for $s\rightarrow-\infty$
\begin{eqnarray}
\nonumber\int_{-\infty}^0\frac{e^{- sz}}{y+z}dy&=&e^{sx}
\int_{-\infty}^z\frac{e^{- su}}{u}du \\
\nonumber& =& e^{sz}\bigg{\{} \bigg{[} \frac{e^{-su}}{-s
u}\bigg{]}_{-\infty}^{z}-\int_{-\infty}^{z}\frac{e^{-su}}{su^2}~du\bigg{\}}\\
\nonumber&=&\frac{1}{(-z s)}\bigg{[}1+\frac{1}{(- z s)}+\frac{2!}{(-
z
s)^2}+\cdots+\frac{n!}{(- z s)^n} +r_n(-z s )\bigg{]}\\
\end{eqnarray}
where the last result  was obtained by integral part by part
repetitions, $z$ can be a complex number, and
\begin{equation}
r_n(z)=(n+1)!ze^{-z}\int_{-\infty}^z\frac{e^{t}}{t^{n+2}}dt.
\end{equation}
and we have \cite{lebsil}
\begin{equation}
|r_n(z)|\leq\frac{(n+1)!}{|z|^{n+1}}.\label{e1}
\end{equation}
Thus, by using the above approximation in the equations (\ref{t5})
and (\ref{eq}) for $s\rightarrow -\infty$ we obtain an estimate of
the survival probability:
\begin{equation}\label{res}
\int_0^{\infty}\int_{-\infty}^{+\infty}
|\mathcal{P}_s\rho_{11}(\nu,E)|^2 d\nu dE \leq
\frac{\gamma^2}{4}[\frac{h(\gamma,\widetilde{\omega}_1)}{\gamma^4s^4}
+ e^{\gamma s }h_1(s,\gamma,\widetilde{\omega}_1)].
\end{equation}
where:
\begin{eqnarray}
\nonumber h(\gamma,\widetilde{\omega}_1)&=&(\frac{256}{\pi\gamma^2
})[\frac{7\pi}{64}+\frac{7}{32}\arctan\frac{2\widetilde{\omega}_1}{\gamma}
-\frac{1}{12}\sin^3(2\arctan\frac{2\widetilde{\omega}_1}{\gamma})
+\frac{1}{4}\sin(2\arctan\frac{2\widetilde{\omega}_1}{\gamma})\\ &
-& \frac{1}{16}\sin(4\arctan\frac{2\widetilde{\omega}_1}{\gamma})
+\frac{1}{256}\sin(8\arctan\frac{2\widetilde{\omega}_1}{\gamma})]
\end{eqnarray}
and
\begin{equation}
h_1(s,\gamma,\widetilde{\omega}_1)=2[\frac{\pi
}{\gamma}\arctan\frac{2\widetilde{\omega}_1}{\gamma}+\frac{\gamma
\sin(2\widetilde{\omega}_1s)-2\widetilde{\omega}_1\cos(2
\widetilde{\omega}_1s)}{s(\widetilde{\omega}_1^2+\frac{\gamma^2}{4})}]
\end{equation}
Here we have an algebraically decreasing function and an
exponentially decreasing multiplied by the oscillating functions.


\section{Conclusion}

We have shown that the pure initial state
$\rho(t)=|\psi_t><\psi_t|$, decomposes into decaying state and a
background, $\rho(t)\rightarrow \mathcal{P}_0\rho(t)+
(1-\mathcal{P}_0\rho(t))$. In the other hand, our result shows that
the survival probability is decreasing for long time exponentially
and algebraically, i.e. we do not have a Zeno effect
\cite{karpov1,karpov2} for our survival probability.

Recently, we have studied 2-level Friedrichs model with weak
coupling interaction constants for a decay phenomena in the Hilbert
space for kaonic system \cite{cds1,cds2}. In future, we shall
consider 2-level or $n$-level Friedrichs by using time
super-operator in the Liouville space to study in order an
irreversible decay description.

\end{document}